\documentclass[aps,prd,10pt,a4paper,superscriptaddress,preprintnumbers,showpacs,notitlepage,eqsecnum,nofootinbib]{revtex4-1}

\usepackage[english]{babel}
\usepackage{graphicx}
\usepackage{amsmath}
\usepackage{xcolor}

\def\square{\mathchoice\sqr54\sqr54\sqr{2.1}3\sqr{1.5}3}
\def\sqr#1#2{{\vcenter{\vbox{\hrule height.#2pt\hbox{\vrule
width.#2pt height#1pt \kern#1pt\vrule width.#2pt}\hrule height.#2pt}}}}
\def\square{\mathchoice\sqr54\sqr54\sqr{2.1}3\sqr{1.5}3}

\begin{document}

\title{Einstein-Katz action, variational principle, Noether charges
\\
and the thermodynamics of AdS-black holes\\
}
\author{Andr\'{e}s Anabal\'{o}n,}

\affiliation{Departamento de Ciencias, Facultad de Artes Liberales y
Facultad de Ingenier\'{i}a y Ciencias, Universidad Adolfo Ib\'{a}\~{n}ez, Vi\~{n}a del Mar, Chile.}

\author{Nathalie Deruelle and F\'elix-Louis Juli\'e}

\affiliation{APC, Universit´e Paris Diderot,\\
CNRS, CEA, Observatoire de Paris, Sorbonne Paris Cit\'e\\
 10, rue Alice Domon et L´eonie Duquet, F-75205 Paris CEDEX 13, France.}

\date{Tuesday July 19th, 2016}

\begin{abstract}
In this paper we describe 4-dimensional gravity coupled to scalar and Maxwell fields by the Einstein-Katz action, that is, the covariant version of the ``Gamma-Gamma $-$ Gamma-Gamma"  part of the Hilbert action supplemented by the divergence of a generalized ``Katz vector". We consider static  solutions of Einstein's equations, parametrized by some integration constants, which describe an ensemble of asymptotically AdS black holes. Instead of the usual Dirichlet boundary conditions, which aim at singling out a specific solution within the ensemble, we impose that the variation of the action vanishes on shell for the broadest possible class of solutions. We will see that, when a long-range scalar ``hair" is present, only sub-families of the solutions  can obey that criterion. The Katz-Bicak-Lynden-Bell (``KBL") superpotential built on this (generalized) vector will then give straightforwardly the Noether charges associated with the spacetime symmetries (that is, in the static case, the mass). Computing the action on shell, we will see next that the solutions which obey the imposed variational principle, and with Noether charges given by the KBL superpotential, satisfy the Gibbs relation, the Katz vectors playing the role of  ``counterterms". Finally, we show on the specific example of dyonic black holes that the sub-class selected by our variational principle satisfies the first law of thermodynamics when their mass is defined by the KBL superpotential.
\end{abstract}

\maketitle

\section{The Einstein-Katz action coupled to scalar and Maxwell fields}

Let us consider the following (``Einstein frame")  action $I$, functional of the fields $g^{\mu\nu}(x^\rho)$, $A_\mu(x^\rho)$ and $\phi(x^\rho)$,  with greek indices running from $0$ to $3$ and coordinates $x^\rho\equiv\{t,r,\theta,\varphi\}$~:
\begin{equation}
2\kappa\, I=
\int \! d^4x\sqrt{-g}\left(R-{1\over2}(\partial\phi)^2-{U(\phi)\over\ell^2}-{1\over4}A(\phi)F^2\right)-\int \! d^4x\sqrt{-\bar g}\left(\bar R+{6\over\ell^2}\right)+\int \! d^4x\,\partial_\mu(\hat k_{\rm K}^\mu+\hat k_{\rm S}^\mu)\,.\label{action}
\end{equation}
Here  $\kappa\equiv8\pi $ (we have set $G=c=1$) ; $\ell$ has the dimension of a length~;  $d^4x\equiv dt\, dr\,d\theta\,d\varphi$~;  $R$ is the scalar curvature of the  metric $g_{\mu\nu}$, with inverse $g^{\mu\nu}$, determinant $g$ and signature $(-,+,+,+)$~;  $(\partial\phi)^2\equiv\partial^\mu\phi\,\partial_\mu\phi\equiv g^{\mu\nu}\partial_\mu\phi\,\partial_\nu\phi$ ; $F^2\equiv F_{\mu\nu}F^{\mu\nu}$ with $F_{\mu\nu}\equiv\partial_\mu A_\nu-\partial_\nu A_\mu$. To keep our presentation simple, we will assume (for reasons explained later) that the expansions for small  $\phi$ of the potential $U(\phi)$ and the coupling function $A(\phi)$  are
 \begin{equation} 
U(\phi)=-6-\phi^2+{\cal{O}}(\phi^4)\quad,\quad A(\phi)=1+{\cal O}(\phi)\,.\label{phi-potential}
\end{equation}

In the second term of (\ref{action}) $\bar g_{\mu\nu}$ and $\bar R=-12\ell^{-2}$ are the metric and scalar curvature of a regularizing anti-de Sitter (AdS) background and all overlined quantities are built with $\bar g_{\mu\nu}$. That background is chosen in order that the total action vanishes for a global AdS spacetime, that is when $g_{\mu\nu}=\bar g_{\mu\nu}$, $\phi=0$, $A_\mu=0$.

The last term is  the integral of a divergence and  will not contribute to the field equations (a hat means multiplication by $\sqrt{-g}$ as in $\hat k^\mu\equiv \sqrt{-g}\, k^\mu$)~; $k_{\rm K}^\mu$ is the ``Katz vector"  (see \cite{Katz85} and, e.g., \cite{KBLB}, for short reviews of its virtues)~:
 \begin{equation} k_{\rm K}^\mu\equiv-(g^{\nu\rho}\Delta^\mu_{\nu\rho}-g^{\mu\nu}\Delta^\rho_{\nu\rho})\quad\hbox{where}\quad \Delta^\mu_{\nu\rho}\equiv\Gamma^\mu_{\nu\rho}-\overline\Gamma^\mu_{\nu\rho}\,,\label{katzvector}
 \end{equation}
 $\Gamma^\mu_{\nu\rho}$ and $\bar\Gamma^\mu_{\nu\rho}$ being respectively the Christoffel symbols associated to the dynamical metric $g_{\mu\nu}$ and the background AdS metric $\bar g_{\mu\nu}$. Again, this vector vanishes when  $g_{\mu\nu}=\bar g_{\mu\nu}$.\footnote{The gravity part of the action can be rewritten as (see, e.g. \cite{KBLB})
 $$
\int\!d^4x\,(\hat R-\overline{\hat R}+\partial_\mu \hat{k}^\mu_{\rm K})=\int\!d^4x\,[\hat g^{\mu\rho}(\Delta^\lambda_{\mu\sigma}\Delta^\sigma_{\rho\lambda}-\Delta^\sigma_{\mu\rho}\Delta^\lambda_{\sigma\lambda})+(\hat g^{\mu\nu}-\overline{\hat g^{\mu\nu}})\bar R_{\mu\nu}]\,,
$$
 hence its name ``Einstein-Katz" action. The Katz vector is closely related to the Gibbons-Hawking-York boundary term : in Gaussian normal coordinates where the metric reads $ds^2=\epsilon dw^2+\gamma_{ij}dx^idx^j$ with $\gamma$ the determinant of the induced metric $\gamma_{ij}$, the GHY boundary term is $2\epsilon\!\int K\sqrt{|\gamma|}d^3x$ where $K\equiv \gamma^{ij}K_{ij}$ is the trace of the extrinsic curvature $K_{ij}={1\over2}\partial_w\gamma_{ij}$, whereas the Katz boundary term reads $\int k^w\sqrt{|\gamma|}d^3x$ with $k^w=2\epsilon\! \left[ K-\frac{1}{2}\bar K_{ij}(h^{ij}+\bar{h}^{ij})\right]$.\\} The divergence of the vector $k^\mu_{\rm S}$ is a new, ``scalar", contribution to the action~:
 \begin{equation} 
k^\mu_{\rm S}\equiv f(\phi)\partial^\mu\phi\label{Svector}
\end{equation}
where the appropriate function $f(\phi)$ will turn out to be
 \begin{equation} 
f(\phi)={\phi\over2}(1+C\phi)\label{ffunction}
\end{equation}
with $C$ a dimensionless constant (which may depend on the parameters entering the potential $U(\phi)$).\footnote{Other divergences of vectors involving $A_\mu$ can be added to the Lagrangian, such as $\partial_\mu (B(\phi)\hat F^{\mu\nu}A_\nu)$~; they will not be required in the examples we shall consider below.\\}

\section{The conditions for an extremal action}

The variation of the action (\ref{action}) under a deformation of the metric, $g^{\mu\nu}\to g^{\mu\nu}+\delta g^{\mu\nu}$ is~:
\begin{equation}
2\kappa\,\delta_{(g)} I=\int\! d^4x\sqrt{-g}\left(G_{\mu\nu}-{1\over2}T_{\mu\nu}\right)\delta g^{\mu\nu}+\int\! d^4x\,\partial_\mu\hat V^\mu_{(g)}\label{metricvariation}
\end{equation}
where $G_{\mu\nu}$ is the Einstein tensor,  where $T_{\mu\nu}$ is the stress-energy tensor of the scalar and Maxwell fields,
\begin{equation}
T_{\mu\nu}\equiv\partial_\mu\phi\partial_\nu\phi-g_{\mu\nu}\left({1\over2}(\partial\phi)^2+\ell^{-2}U(\phi)\right) +A(\phi)\left(F_\mu^{\ \rho} F_{\nu\rho}-{1\over4}g_{\mu\nu}F^2\right)\,,\label{stressEnergy}
\end{equation}
and where 
\begin{equation}
\hat V^\mu_{(g)}\equiv\Delta^\rho_{\nu\rho}\delta\hat g^{\mu\nu}-\Delta^\mu_{\nu\rho}\delta\hat g^{\nu\rho}+f\partial_\nu\phi\,\delta\hat g^{\mu\nu}\,. \end{equation}
(The role of the Katz-vector $k^\mu_{\rm K}$ is to eliminate all terms in $\delta(\partial_\rho g_{\mu\nu})$ generated by the variation of $R$, see \cite{Katz85}.)

Similarly the variation of $I$ under a deformation of the scalar field, $\phi\to\phi+\delta\phi$, is
\begin{equation}
2\kappa\,\delta_{(\phi)} I=\int\! d^4x\sqrt{-g}\left(\square\,\phi-{1\over\ell^{2}}{dU\over d\phi}-{1\over4}{dA\over d\phi}F^2\right)\delta\phi
+\int\! d^4x\,\partial_\mu\hat V^\mu_{(\phi)}\ \hbox{with}\ \hat V^\mu_{(\phi)}=\hat\partial^\mu\phi\left({d\!f\over d\phi}-1\right)\delta\phi+f\hat g^{\mu\nu}\delta(\partial_\nu\phi)\,.\label{phivariation}
\end{equation}
Finally the variation of $I$ with respect to the gauge field $A_\mu$ is
\begin{equation}
2\kappa\,\delta_{(A)} I=\int\! d^4x\sqrt{-g}\,D_\mu\left(A(\phi)F^{\mu\nu}\right)\delta A_\nu+\int\! d^4x\,\partial_\mu\hat V^\mu_{(A)} \quad\hbox{with}\quad \hat V^\mu_{(A)}=-A(\phi)\hat F^{\mu\nu}\delta A_\nu\,.\label{Avariation}
\end{equation}

As a first condition to extremize the action, we impose that the fields obey the Einstein, Klein-Gordon and Maxwell equations~:
\begin{equation}
G_{\mu\nu}={1\over2}T_{\mu\nu}\quad,\quad \square\,\phi-{1\over\ell^{2}}{dU\over d\phi}-{1\over4}{dA\over d\phi}F^2=0 \quad,\quad D_\mu\left(A(\phi)F^{\mu\nu}\right)=0\,,\label{eom}
\end{equation}
(the Klein-Gordon equation, say, following from the others thanks to the Bianchi identity).\\

 A second condition to guarantee that the variation of the action vanishes is that the vector densities $\hat{V}^\mu$  be zero on the boundary, which is composed of 2 spacelike hypersurfaces $\Sigma_1$ and $\Sigma_2$ (whose equations, in adapted coordinates, are $x^0 = t_1$ and $x^0 = t_2$, with $t_2 - t_1 = \beta$), together with a timelike cylinder formed by the piling of the 2-spheres at infinity with equations $x^0 = const$, $x^1 = r\to\infty$.

  A way to implement that is 
 to impose Dirichlet conditions, that is that the fields are fixed on the boundary a priori~: $\delta g^{\mu\nu}=0$, $\delta\phi=0$ and $\delta A_\mu=0$ on the boundary. As in point-mechanics, such conditions are aimed at selecting a particular solution of the field equations, that is at fixing to arbitrary, but specific, values  its integration constants  (to wit the mass and charge parameters which play the role of initial and final ``positions"). If such Dirichlet conditions are imposed, then the function $f(\phi)$ which appears in $\hat V^\mu_{(\phi)}$ in (\ref{phivariation}) must be absent and the boundary term in the action (\ref{action}) then reduces to the Katz-vector.

Those are not the conditions we will choose here. Since, rather than specific,  we want to consider {\sl families} of (black hole) solutions in order to study their thermodynamics we shall impose that the vectorial densities $\hat V^\mu_{(g)}$, $\hat V^\mu_{(\phi)}$ , $\hat V^\mu_{(A)}$ vanish on the boundary when evaluated on shell, that is upon variation of the integration constants that characterize the solutions (their mass, possibly angular momentum, gauge and scalar parameters). 

As we shall see below in the case of static, asymptotically AdS, ``hairy" black holes, the introduction of these Katz vectors together with the proposed variational principle will simplify some calculations and may
shed a new light on the thermodynamics of AdS black holes.

\section{Constraints on static, spherically symmetric\\ and asymptotically AdS solutions\\
imposed by the variational principle}

Consider a  static and spherically symmetric configuration of the fields. In Schwarzschild-Droste coordinates, the metric and fields read :
\begin{equation}
ds^2=-h(r)\,dt^2+{dr^2\over \tilde h(r)}+r^2(d\theta^2+\sin^2\!\theta\, d\varphi^2)\quad,\quad \phi=\phi(r)\quad\hbox{and}\quad A_\mu=(-V(r),0,0,4P\cos\theta)\,,\label{metric0}
 \end{equation}
 (where the $A_\phi$ component of $A_\mu$, with $P$ a constant with dimension of a length,  allows for a magnetic charge $P$).
 \\
 
  The Klein-Gordon and Maxwell equations, see (\ref{eom}), then read (a prime denoting derivation wrt $r$)
   \begin{equation}
 \phi'\tilde h \left[\ln\left(\phi'r^2\sqrt{h\tilde h}\right)\right]'=\ell^{-2}{dU\over d\phi} +{1\over4}{dA\over d\phi}F^2\quad,\quad \left(r^2\sqrt{\tilde h/h}\,AV'\right)'=0 \label{KG}
   \end{equation} 
and the $t$-$t$, together with the difference between the $t$-$t$ and $r$-$r$ components of Einstein's equations close the system~:
\begin{equation}
r\tilde h'+\tilde h-1+
{r^2\over4}\left[\tilde h\,\phi'^2+{2U\over\ell^2}+A(\phi)\left({16P^2\over r^4}+{\tilde h \over h}(V')^2\right)\right]=0\quad,\quad \phi'^2={2\over r}\left(\!\ln{h\over\tilde h}\right)'\,.\label{Eeq}
 \end{equation} 
   
 Suppose now that $A(\phi)=1+{\cal O}(\phi)$ and $U(\phi)=-6-\phi^2+{\cal O}(\phi^4)$, as assumed in (\ref{phi-potential}).\footnote{A generic behaviour of the scalar potential, $U(\phi)=-6+\mu^2\phi^2+\lambda\phi^3+\cdots$ , changes the fall-off of the scalar field and renders the analysis more involved, see, e.g., \cite{Henneauxetal}, \cite{Andresetal15}, \cite{luPopeWen}, and references therein.} Solving iteratively the system (\ref{KG}-\ref{Eeq}), one finds that the solution behaves  asymptotically as, see, e.g.,  \cite{Popeetal} (the Coulomb potential $V(r)$ is defined up to a constant which is chosen so that it vanishes at infinity)~:
  \begin{equation}
\phi(r)={\phi_1\over r}+{\phi_2\over r^2}+{\cal O}(1/r^3)\quad,\quad V(r)={4Q\over r}+{\cal O}(1/r^2)\label{solphi}
   \end{equation}
 \begin{equation}
h(r)=\ell^{-2}r^2+1-{2m_g\over r}+{\cal O}(1/r^2)\quad,\quad \tilde h(r)=\ell^{-2}r^2 +1+\alpha^2-{2m_i\over r}+{\cal O}(1/r^2)\,.\label{metric}
 \end{equation}
with~:
  \begin{equation}
\alpha={\phi_1\over2\ell} \quad\hbox{and}\quad  m_i=m_g-{\phi_1\phi_2\over3\ell^{2}}\,.\label{solmetric}
\end{equation}
Once the coupling function $A(\phi)$ and the potential $U(\phi)$ are explicitly given, all subsequent coefficients in the $1/r$ expansion of the metric, as well as of the scalar and Maxwell fields, can be expressed in function of the 5 integration constants, $\phi_1$, $\phi_2$, $m_g$, together with the magnetic and electric charges $P$ and $Q$. (Of course, specific black hole solutions may depend on fewer parameters, as we shall see in Section VI.)
\\

The solution (\ref{solphi}-\ref{metric}-\ref{solmetric}) of the field equations will extremize the action if the radial components of $\hat V^\mu_{(g)}$,  $\hat V^\mu_{(\phi)}$ and $\hat V^\mu_{(A)}$, defined in (\ref{metricvariation}-\ref{phivariation}-\ref{Avariation}),  vanish on the boundary (their $t$,   $\theta$ and $\varphi$ components  are zero because the solution is static and spherically symmetric.)

To evaluate  $\hat V^r_{(g)}$  we write the AdS background metric in the same, Schwarzschild-Droste, coordinates, that is~:
\begin{equation} 
 d\bar s^2=-\bar h(r)dt^2+{dr^2\over \bar{\tilde h}(r)}+r^2(d\theta^2+\sin^2\theta d\phi^2)\quad\hbox{with}\quad  \bar h=\bar{\tilde h}=\ell^{-2}r^2+1\,.\label{backgroundAdS}
   \end{equation} 
  
To evaluate $\hat V^r_{(\phi)}$, we take the following ansatz  for $f(\phi)$~:
\begin{equation}
f(\phi)=A +B\phi+C{\phi^2\over2}\cdots
\end{equation}
where the constants $A$, $B$ and $C$ will be chosen in order that $\hat V^r_{(g)}$ and $\hat V^r_{(\phi)}$ vanish on the boundary.   (The Maxwellian boundary term, $\hat V^r_{(A)}$, turns out to vanish at spatial infinity because $F^{tr}=-F^r_t/h$ contains the extra factor $h^{-1}={\cal{O}}(1/r^2)$ as compared to $\partial^r\phi$.)
   
The calculation of  $\hat V^r_{(g)}$ and $\hat V^r_{(\phi)}$  uses (omitting the irrelevant factor $\sin\theta$ in $\sqrt{-g}=r^2\sin \theta\sqrt{h/\tilde h}\,$) :
\begin{equation}
\hat g^{tt}=-{r^2\over\sqrt{h\tilde h}}\quad,\quad\hat g^{\theta\theta}=\sqrt{{h\over\tilde h}}\quad,\quad\hat g^{rr}=r^2\sqrt{h\tilde h}\label{calcul1}
\end{equation}
\begin{equation}
 \Delta^r_{tt}={\tilde h h'-\overline{\tilde{h} h'}\over2}\quad,\quad \Delta^r_{\theta\theta}=-r(\tilde h-\overline{\tilde h})\quad,\quad \Delta^t_{rt}={h'\over2h}-\overline{h'\over2\bar h}\label{calcul2}
 \end{equation}
\begin{equation}
\delta h=-2{\delta m_g\over r}+{\cal{O}}(1/r^2)\quad,\quad \delta\tilde h=2\alpha\delta\alpha-2{\delta m_i\over r}+{\cal{O}}(1/r^2)\quad,\quad \delta\phi={\delta\phi_1\over r}+{\delta\phi_2\over r^2}+{\cal{O}}(1/r^3)\,.
\end{equation}
(Of course $\kappa$ and $\ell$ are not varied as they are ``universal" constants entering the action.)  The result is -- using (\ref{solphi}) and (\ref{metric}) but without the need to impose (\ref{solmetric}) :
\begin{equation}
\hat V^r_{(g)}=-A\phi_1\alpha\delta\alpha+{\cal{O}}(1/r)\,,\label{Vrg}
\end{equation}
\begin{equation}\hat V^r_{(\phi)}=r^2\,\ell^{-2}A\,\delta\phi_1+r\,\ell^{-2}(2A\,\delta\phi_2+\phi_1\delta\phi_1(2B-1))+{\cal{O}}(1)\,.\label{Vrphi}
\end{equation}
We hence first recover from (\ref{Vrg}) the well-known result that, in the absence of scalar field (and thus in the absence of $f$), the variation of the action on shell is zero $\forall$ $\delta m$ (and $\forall$ $\delta Q$ and $\delta P$).
To ensure now that $\hat V^r_{(g)}$ and $\hat V^r_{(\phi)}$ vanish at spatial infinity {\it for the largest possible family of solutions}, i.e. when neither $\phi_1$ nor $\phi_2$ are given a priori (or, else, when neither $\delta\phi_1$ nor $\delta\phi_2$ are imposed to be a priori zero ``\`a la" Dirichlet), we must have~:

({\it i}) first, see (\ref{Vrphi}),  that  $A=0$  (which implies $\hat V^r_{(g)}={\cal O}(1/r)$), $B=1/2$ and, thus, as anticipated in (\ref{ffunction})~:
\begin{equation}
f(\phi)={\phi\over2}(1+C\phi)+{\cal{O}}(\phi^3)\,,\label{functionf}
\end{equation}
where $C$ is an arbitrary constant,\\

({\it ii}) and, these conditions being fulfilled, we must have, second, that the ${\cal{O}}(1)$-term of $\hat V^r_{(\phi)}$, which reduces then to 
\begin{equation}
\hat V^r_{(\phi)}=-{1\over2\ell^2}\left(\phi_1\delta\phi_2-(\phi_2-3C\phi_1^2)\delta\phi_1\right)+{\cal O}(1/r)\,,
\end{equation}
vanishes as well at spatial infinity, $\hat V^r_{(\phi)}=0$, which implies that $\phi_1$ and $\phi_2$ cannot in fact be entirely free but must be related as
\begin{equation} 
\phi_2=-3C\phi_1^2+D\ell\phi_1\label{phi2condition}
\end{equation}
where $D$ is another arbitrary number. 

Therefore $\phi_1$ and $\phi_2$ can never be {\it independent} constants of integration of the Einstein equations.

The numbers $C$ and $D$ are to be seen as ``universal", that is determined by the {\it theory}, and not, contrarily to $\phi_1$, $m_g$, $P$ or $Q$, by its {\it solutions}. They do not have the same status however~: $C$ enters the action through the boundary term $k^\mu_{\rm S}=f(\phi)\partial^\mu\phi$, and should be put on the same footing as $\kappa$ and $\ell$, whereas $D$ does not. One may therefore argue that $D$ must be taken to be zero, in which case the scalar field asymptotic behaviour preserves the AdS symmetry, see e.g. \cite{Henneauxetal} or \cite{Peresetal}.

As is well known, scalar fields saturating the
Breitenlohner-Freedman bound, $\mu^2\geq -9/4$ (see footnote 4), ensure the
linear stability of AdS when $\phi_1=0$ \cite{Breitenlohner:1982bm}. When $\phi_1 \neq 0$ a detailed analysis
shows that the condition for linear stability, in the case when the
scalar field mass is $\mu^2=-2$, is $\frac{1}{\ell}\frac{d\phi_2}{d\phi_1}\mid_{\phi_1=0}\geq
-2/\pi$. Hence  $D$ is constrained by stability
as $D \geq -2/\pi$ \cite{Anabalon:2015vda}.

To summarize, by introducing an adequate function  $f$, we have built an action  whose variation is zero on shell for the broadest possible family of solutions. The fact that this family has to be restricted by the constraint (\ref{phi2condition})  is the first result of this paper.

\section{Mass as a Noether charge}

The Katz-Bicak-Lynden-Bell superpotential is the ``covariantization" of Freud's superpotential and hence is based on a covariantization of Einstein's pseudo-tensor, see \cite{Katz85}  and \cite{KBLB} or \cite{JuliaSilva}. That Lagrangian definition of Noether charges has been successfully applied to various spacetimes over the years, including D-dimensional, rotating, asymptotically flat or AdS  black hole solutions of pure Einstein or Einstein-Gauss-Bonnet equations, see \cite{DKO}, \cite{DKkerr}, \cite{DM} \cite{Petrov}~; for  a comparison with other approaches and, in particular, the Ashtekar-Magon-Das definition of the mass, see \cite{DKconformal}.\\

If the gravity theory is Einstein's, the following identity, see e.g. \cite{DKO}, is obtained by exploiting the invariance under diffeomorphisms $x^\rho\to x^\rho+\xi^\rho$ of the action $I_g=\int d^4x(\hat R-\bar{\hat R})+\int d^4x\,\partial_\mu(\sqrt{-g}\,k^\mu)$ (matter fields do not play any role)~:
\begin{equation}
\int\!d^4x\,\partial_{\mu\rho}\hat J^{[\mu\rho]}\equiv0\quad \hbox{where}\quad
 \kappa \hat J^{[\mu\rho]}=D^{[\mu}\hat\xi^{\rho]}-\overline{D^{[\mu}\hat\xi^{\rho]}}+\xi^{[\mu}\hat k^{\rho]}\quad\hbox{with}\quad k^{\rho}= k_{\rm K}^{\rho}+ k_{\rm S}^{\rho}\,.\label{superpotential}
 \end{equation}
 The ``superpotential" $J^{[\mu\nu]}$ was first introduced by Katz in the case of a flat background, see \cite{Katz85}, and then extended by Katz, Bicak and Lynden-Bell to an arbitrary background in \cite{KBLB} and thus is usually called the ``KBL superpotential", see also \cite{JuliaSilva}~; brackets denote antisymmetrization (e.g. $\xi^{[\mu} k^{\rho]}\equiv(\xi^\mu k^{\rho}-\xi^\rho k^\mu)/2$)~; the overline over $D^{[\mu}\hat\xi^{\rho]}$ means that it is evaluated using the background metric $\bar g_{\mu\nu}$ ; finally $k_{\rm K}^{\rho}$ and $k_{\rm S}^{\rho}$ are given in (\ref{katzvector}) and (\ref{Svector}).
One  recognizes in the first two terms half of the (regularized) Komar superpotential. The role of the Katz vector $k_{\rm K}^{\rho}$ is, in particular, to correct the ``anomalous factor 2" of Komar's formulae, see \cite{Katz85}. The contribution of the vector $k_{\rm S}^{\rho}$ is new, and due to the presence of a long-range scalar field.

Equation (\ref{superpotential}) yields conservation laws. If spacetime is stationary they read (in adapted coordinates)~:
\begin{equation}
\lim_{r\to\infty}\int_S\!d\theta\, d\phi\, \hat J^{[0r]}=Const.
\end{equation}
 where $S$ is the 2-sphere at spatial infinity with equation $r\to\infty$.
 \\
 
   If one is interested in defining the mass $M$, $\xi^\mu$ is taken to be the Killing vector for time translations, $\xi^\mu=(1,0,0,0)$, and the constant is identified with $-M$.  
\\

Consider now the asymptotically AdS configurations (\ref{solphi}) and (\ref{metric}) for the fields and the metric. Using (\ref{calcul1}-\ref{calcul2}) and choosing the function $f(\phi)$ given in (\ref{functionf}) a short calculation gives
\begin{equation}
M={1\over8}(\ell^{-2}\phi_1^2-4\alpha^2)r+\left[m_i+{1\over8}\ell^{-2}\phi_1(3\phi_2+C\phi_1^2)\right]+{\cal{O}}(1/r)\,.
\end{equation}
Imposing now that ({\it i}) the configuration (\ref{solphi}-\ref{metric}) indeed solves Einstein's equation, that is, imposing conditions (\ref{solmetric}), and ({\it ii}) that the solution extremizes the action, that is, imposing condition (\ref{phi2condition}), then yields
\begin{equation}
M=m_g+D{\phi_1^2\over24\ell}\,.\label{mass}
\end{equation}
 Hence, the mass of the solutions of the field equations which extremize the action is given by (\ref{mass}). When $D\neq0$ that extends the results obtained in ref \cite{Peresetal}. 
 The expression (\ref{mass}) for the mass of asymptotically AdS hairy black holes is the second result of this paper.
 
 \section{Euclidean action and Gibbs' relation}
 
 The value of the action on shell plays an important role in the study of  black hole thermodynamics. We compute it here and show that it can be identified to a Gibbs potential, the Katz vectors playing the role of the counterterms introduced in \cite{Henningson} and used in, e.g., \cite{Andresetal15} \cite{Popeetal}, and references therein.
 
Let us first treat the bulk part of the action (\ref{action}), that is
\begin{equation}
2\kappa I_{\rm bulk}\equiv\int\!d^4x\sqrt{-g}\left(R-{1\over2}(\partial\phi)^2-{U(\phi)\over\ell^2}-{1\over4}A(\phi)F^2\right)\,.\label{bulkaction}
\end{equation}
$I_{\rm bulk}$ reduces to a boundary term on shell. Indeed the Gauss-Codazzi equations (or a direct manipulation of the Einstein equations (\ref{KG}) and (\ref{Eeq})), reduce, in the case of a static, spherically symmetric metric, written in Schwarzschild coordinates as in (\ref{metric0}), to :
$$\hat R +2\hat G^t_t=-\left(r^2h'\sqrt{\tilde h/h}\right)' \,.\eqno(5.2)$$
Now the $t$-$t$ equation of motion, after adding $\hat R$ to both sides reads $\hat R +2\hat G^t_t=\hat R+\hat T^t_t$, that is, see (\ref{stressEnergy})
\begin{equation}
\hat R+2\hat G^t_t=\sqrt{-g}\left(R-{1\over2}(\partial\phi)^2-{U(\phi)\over\ell^2}-{1\over4}A(\phi)F^2\right)+A_0\,\partial_\mu(A\hat F^{0\mu})-(A\hat F^{0r}A_0)'\label{Ttt}
\end{equation}
where one recognizes in the first term the integrand of $I_{\rm bulk}$. As for Maxwell's equations they are $\partial_\mu(A\hat F^{\mu\nu})=0$, so that the second term on the r.h.s. of (\ref{Ttt}) is zero on shell~; as for  their first integral it gives $A\hat F^{0r}=4Q$, which simplifies the last term (recalling that $A_0=-V$ and that $V=4Q/r+{\cal O}(1/r^2)$, see (\ref{solphi})). 
Therefore
\begin{equation}
2\kappa I_{\rm bulk}|_{\rm on shell}=-4\pi\beta\left(r^2h'\sqrt{\tilde h/ h}+4QV\right)^{r\to\infty}_{r=r_+}\label{Ibulk}
\end{equation}
where $4\pi$ is the integral over the angles, where $\beta\equiv t_2-t_1$ is the time integral, and where the spatial boundaries of the manifold are taken to be spatial infinity and, as usual, the black hole horizon.

The lower boundary of (\ref{Ibulk}) is evaluated on the (supposedly non-degenerate) horizon where $h=h'_+(r-r_+)+\cdots$, $\tilde h=\tilde h'_+(r-r_+)+\cdots$, so that $4\pi\beta \left(r^2h'\sqrt{\tilde h/ h}\right)|_{r_+}=4\pi\beta r^2_+\sqrt{\tilde h'_+h'_+}$. The temperature $T$ of the black hole, defined as, e.g., $1/2\pi$ the horizon surface gravity, is given by $T=\sqrt{\tilde h'_+h'_+}/(4\pi)$. Its entropy $S$  is defined as one-fourth of its area, that is~: $S=\pi r_+^2$. Finally the time interval $\beta$ is taken to be the inverse of the temperature.
Hence, all in all, we get the standard result, see, e.g., \cite{Popeetal}~:
\begin{equation}
4\pi\beta\left(r^2h'\sqrt{\tilde h/ h}+4QV\right)_{r_+}=16\pi (S+\beta Q\Phi_Q)\,,\label{lowerbound}
\end{equation}
where $\Phi_Q$ is the value of the electric potential on the horizon. 

Inserting now the asymptotic solution (\ref{solphi}), (\ref{metric}) and (\ref{solmetric})  in (\ref{Ibulk}), the upper boundary  reads
\begin{equation}
\lim_{r\to\infty}-4\pi\beta\left(r^2h'\sqrt{\tilde h/ h}+4QV\right)=-{8\pi\beta \over\ell^2}r^3-4\pi\beta\,{\phi_1^2\over4\ell^2}\,r+4\pi\beta\left(-2m_g-{2\over3}{\phi_1\phi_2\over\ell^2}\right)\,.\label{upperbound}
\end{equation}
As one can see, and is well-known, that upper boundary term  diverges, first because the metric is asymptotically AdS, and, second, because of the presence of a long range scalar field (as for the electromagnetic field, it does not contribute). Usually those divergences are eliminated by means of counterterms added to the GHY surface term, that is boundary terms which depend on its curvature, see e.g.  \cite{Andresetal15} or \cite{Popeetal}. The addition of such terms does not prove necessary in the present framework, as the divergences are cancelled by the background bulk action and the Katz boundary terms, as we now show in detail.

To the bulk action on shell, that is the sum of (\ref{upperbound}) and (\ref{lowerbound}), we must indeed now add the contributions on shell of the background action, $2\kappa \bar I=-\int \! d^4x\sqrt{-\bar g}\left(\bar R+{6\over\ell^2}\right)$, and the generalized Katz term, $2\kappa I_{\rm b}=\int \! d^4x\,\partial_\mu(\hat k_{\rm K}^\mu+\hat k_{\rm S}^\mu)$. The background being AdS in Schwarzschild coordinates, see (\ref{backgroundAdS}), its action on shell is trivially given by
\begin{equation}
2\kappa \bar I|_{\rm on shell}={6\over\ell^2}4\pi\beta\int_0^rdr\,r^2={8\pi\beta\over\ell^2}r^3\label{Ibarbulk}
\end{equation}
where the notation makes it clear that, AdS spacetime being regular, the lower bound is taken to be $r=0$. That term cancels the first (divergent) term in (\ref{upperbound}).

As for the computation of $2\kappa I_{\rm b}=\int \! d^4x\,\partial_\mu(\hat k_{\rm K}^\mu+\hat k_{\rm S}^\mu)$, it proceeds as follows (after integration on time and on the angles and omitting the factor $\sin\theta$ in $\sqrt{-g}$)~:
\begin{equation}
2\kappa I_{\rm b}=4\pi\beta(\hat k_{\rm K}^r+\hat k_{\rm S}^r)|^{r\to\infty}\label{Iboundary}
\end{equation}
where here, as above, the inner boundary is taken to be the center of the manifold where the vectors, by symmetry, are taken to vanish. Using (\ref{solphi}) and (\ref{metric}) one has
\begin{equation}
\hat k^r_{\rm S}\equiv f(\phi)\hat\partial^r\phi={\phi\over2}(1+C\phi)r^2\sqrt{h\tilde h}\,\phi'=-{\phi_1^2\over2\ell^2}r-{\phi_1\over2\ell^2}(3\phi_2+C\phi_1^2)+{\cal O}(1/r)\,,
\end{equation}
and, using (\ref{calcul1}) and  (\ref{calcul2}) as well as the metrics (\ref{metric}) and (\ref{backgroundAdS})~:
\begin{equation}
\hat k^r_{\rm K}\equiv-(\hat g^{\nu\rho}\Delta^r_{\nu\rho}-\hat g^{r\nu}\Delta^\rho_{\nu\rho})=3\alpha^2r+(4m_g-6m_i)+{\cal O}(1/r)\,.
\end{equation}
Since the e.o.m. impose $\alpha=\phi_1/(2\ell)$ and $m_i=m_g-2\phi_1\phi_2/(3\ell^2)$, see (\ref{eom}), the boundary term finally reads
\begin{equation}
2\kappa I_{\rm b}|_{\rm on shell}=4\pi\beta\,{\phi_1^2\over4\ell^2}\,r+4\pi\beta\left(-2m_g+{\phi_1\phi_2\over2\ell^2}-C{\phi_1^3\over2\ell^2}\right)\,.
\end{equation}
As the detailed calculation presented here makes it clear, both vectors $k^\mu_{\rm K}$ and $k^\mu_{\rm S}$, with the function $f(\phi)$ imposed by our variational principle, play a role in cancelling the second, divergent, term in   (\ref{upperbound}).

Adding (\ref{upperbound}), (\ref{lowerbound}), (\ref{Ibarbulk}) and (\ref{Iboundary}), the action on shell is finite and reads
\begin{equation}
I_{\rm on shell}=S+\beta Q\Phi_Q-\beta\left(m_g+{\phi_1\phi_2\over24\ell^2}+C{\phi_1^3\over8\ell^2}\right)\,.\label{Ionshell}
\end{equation}
The last piece of information we have not used yet is that our variational principle imposes that $\phi_2=-3C\phi_1^2+D\ell\phi_1$, see (\ref{phi2condition}), and that the KBL superpotential defines the mass as $M=m_g+D{\phi_1^2\over24\ell}$, see (\ref{mass}), so that the last term in (\ref{Ionshell}) is nothing but the mass. Hence~:
\begin{equation}
I_{\rm on shell}=S-\beta(M-Q\Phi_Q)\,,\label{Gibbsrelation}
\end{equation}
which is the Gibbs relation when, as usual, one interprets $-I_{\rm on shell}/\beta $ as the black hole Gibbs potential\footnote{The Gibbs potential is related to the Euclidean action $I^E$ by $G=I^E/\beta$, where $I^E=-iI$ and $I$ is the Lorentzian action integrated in imaginary time $t\in [0,-i\beta]$, i.e. $I=-iI_{onshell}$ and thus $I^E=-I_{onshell}$. We note that, since we imposed that the variation of the action on shell be zero, the on-shell action itself does not depend on the integration constants ($m_g$, $\phi_1$, $\phi_2$, $Q$ and $P$) and hence depends only on $\Sigma_1$ and $\Sigma_2$, that is $\beta$.\\} (note the absence of the magnetic charge, as in \cite{Popeetal}, a result which, as noted in  \cite{Popeetal}, can be modified by a proper adjunction to the action of boundary terms involving Maxwell's field, see footnote 2). Consider as an example  the asymptotically AdS black hole solution (with no electric nor magnetic charges) discovered by one of us in \cite{Andressol}~: it obeys the Gibbs relation simply because the asymptotic fall-off of the scalar field is such that $\phi_2=-3C\phi_1^2+D\ell\phi_1$ with $D=0$ and $2C=-\sqrt{\nu^2-1}$, $\nu$ being a parameter entering the definition of the scalar potential $U(\phi)$.

Arriving at (\ref{Gibbsrelation}) by means of the Katz vectors is the third, and main, result of this paper.\footnote{Hence the Katz vectors replace, in the present framework, the sum of the GHY surface term, see footnote 1, and the following counterterms~: $-{1\over\kappa}\int\!d^3x\sqrt{|\gamma|}\left({2\over\ell}+{{\cal R}\ell\over2}\right)$, where $\gamma_{ij}$ is the metric on the boundary and ${\cal R}$ its curvature scalar, augmented by a scalar contribution, taken to be ${1\over6\kappa}\int\!d^3x\sqrt{|\gamma|}\left(\phi\,n^\nu\partial_\nu\phi-{\phi^2\over2\ell}\right)$, see e.g. \cite{Popeetal} and references therein,  and see \cite{Andresetal15} for an alternative proposal.\\}\\

\section{Dyonic black-holes and their thermodynamics}

In {\cite{Popeetal} L\"u et al. (henceforth LPP) considered the following bulk action~:
\begin{equation}
I[g^{\mu\nu},\phi, A_\mu]\equiv\int \! d^4x\sqrt{-g}\left(R-{1\over2}\partial^\mu\phi\,\partial_\mu\phi+6\ell^{-2}\cosh\left({\phi\over\sqrt3}\right)-{1\over4}e^{-\sqrt3\,\phi}F_{\mu\nu}F^{\mu\nu}\right)\label{Popeaction}
\end{equation}
which falls in the class studied above since $U(\phi)\equiv-6\cosh\left({\phi/\sqrt3}\right)=-6-\phi^2+\cdots$ and $A(\phi)\equiv\exp (-\sqrt3\,\phi)=1-\sqrt3\,\phi+\cdots$.

Restricting one's attention, as we did above,  to static, spherically symmetric, asymptotically AdS  solutions of the derived equations of motion, the leading orders in $1/r$ of the metric and the scalar and electromagnetic fields  are given by (\ref{solphi}), (\ref{metric}) and (\ref{solmetric}).  As previoulsy discussed the solutions obtained by solving iteratively the field equations depend on 5 integration constants, $\phi_1$, $\phi_2$, $m_g$, $Q$ and $P$.

Now, L\"u et al. found a remarkable sub-family of black-hole solutions with that  asymptotic behaviour, characterized by 3 parameters only, which can be taken to be  $\phi_1$, $\phi_2$ and $m_g$, $Q$ and $P$ becoming specific functions of $\phi_1$, $\phi_2$ and $m_g$, see Eq. (2.1) of  {\cite{Popeetal} for their explicit expression.\footnote{Note that in \cite{Popeetal} $\rho$, and not $r$, is the Schwarzschild coordinate. Note too that, instead of  $\phi_1$, $\phi_2$ and $m_g$, LPP use 3 other parameters called $\beta_1$, $\beta_2$ and $\mu$ which are implicitely related to $\phi_1$, $\phi_2$ and $m_g$ through their Eq. (2.7-9).}

The horizon $r_+$ of the black hole is the common zero of $g_{tt}$ and $g^{rr}$, that is is such that 
 \begin{equation}
 h[r_+]=0\quad,\quad \tilde h[r_+]=0\,.
 \end{equation}
 Hence $r_+=r_+(\phi_1,\phi_2, m_g)$ ; see LPP eq. (2.10) for its (implicit) expression.

 The electric and magnetic potentials are also easily defined and are also functions of $\phi_1$, $\phi_2$ and $m_g$, see LPP eq. (2.12) for their values ($\Phi_Q$ and $\Phi_P$) on the horizon.

 The temperature $T$  and entropy $S$ of the black-hole are defined as is usual in Einstein's theory and are, as well, known functions of $\phi_1$, $\phi_2$ and $m_g$, see LPP eq. (2.11).
 
 Finally, L\"u et al. calculate the mass of the black hole using the Astekhar-Magnon-Das method which yields 
  \begin{equation}
 M_{\rm LPP}=m_g\,.
 \end{equation}
 (The Hamiltonian-Wald approach, see  \cite{Klemm}, \cite{LiuLu}, \cite{Andresetal}, \cite{valdivia}, yields the same result.)
 \\
 
 With all that information in hand it is an exercise to study the thermodynamics of the black hole.
 The first law is satisfied if 
 \begin{equation}
L_{\rm LPP}\equiv  TdS-(dM_{\rm LPP}-\Phi_P dP-\Phi_Q dQ)\label{thermopb1}
 \end{equation}
vanishes. Expanding  $dS$ etc in $dS=(\partial S/\partial \beta_1)d\beta^1+(\partial S/\partial \beta_2)d\beta^2+(\partial S/\partial \mu)d\mu$ etc, the explicit calculation shows  that $L_{\rm LPP}$  does {\it not} vanish. As shown in \cite{Popeetal}, see also \cite{luPopeWen} and \cite{LiuLu}, one rather has that
  \begin{equation}
L_{\rm LPP}=\frac{1}{12\ell^2}(2\phi_2\,d\phi_1-\phi_1\,d\phi_2)\,.\label{thermopb2}
 \end{equation}
 
Various solutions for that puzzle have been proposed.

L\"u et al. in \cite{Popeetal} rewrite the right-hand-side of (\ref{thermopb2}) as $X dY$ with $X$ and $Y$ being some functions of $\beta_1$ and $\beta_2$ given in LPP eq. (2.14). Despite the fact that $X$ and $Y$ are not uniquely defined, LPP interpret $Y$ as some ``scalar charge" and $X$ as the corresponding scalar potential evaluated on the horizon. That solution to the problem has been criticized in, e.g., \cite{Compere}. It is indeed unclear how a scalar charge can be defined as there is no global symmetry associated to scalar fields.

Cardenas et al.  impose in \cite{valdivia} that the asymptotic scalar field (which behaves as $\phi=\phi_1/r+\phi_2/r^2+\cdots$) preserve the asymptotic AdS symmetries. That imposes, see \cite{Henneauxetal}, that $\phi_2=-3C\phi_1^2$ with $C$ an arbitrary number, see also \cite{LiuLu}. The same result is obtained along a different route in \cite{Andresetal}.\\

We propose here still  another solution to the puzzle : (1) impose that the LPP back hole solutions obey the variational principle advocated in this paper,  hence proper boundary terms have to be added to the bulk action (\ref{Popeaction})~; (2) define the mass ``\`a la" Katz rather than ``\`a la" AMD.\\

Let us analyze  these 2 points :

(1) After adding to the action (\ref{Popeaction})  the boundary terms built from the vectors $k^\mu_{\rm K}$ and $k^\mu_{\rm S}$ --see (\ref{katzvector}) and  (\ref{Svector}) with $f(\phi)$ given in (\ref{functionf})--, and subtracting a regularizing AdS background, the LPP black-hole solutions  extremize the action if $\phi_1$ and $\phi_2$ are not independent but related as, see (\ref{phi2condition})~:
\begin{equation}
\phi_2=-3C\phi_1^2+D\ell\phi_1\,.
\end{equation}

(2) The mass of the LPP black hole solutions satisfying that constraint, as obtained by means of the KBL superpotential, is given by (\ref{mass})~: 
\begin{equation}
M=m_g+D{\phi_1^2\over24\ell}\quad\hbox{that is}\quad M=M_{\rm LPP}+D{\phi_1^2\over24\ell}\,.
\end{equation}

Therefore, we have on one hand : $dM_{\rm LPP}=dM-D\phi_1\,d\phi_1/12\ell$, and, on the other hand : $(2\phi_2\,d\phi_1-\phi_1\,d\phi_2)/(12\ell^2)=D\phi_1\,d\phi_1/12\ell$, so that (\ref{thermopb1}) and (\ref{thermopb2}) yield
\begin{equation}
dM=TdS+\Phi_P dP+\Phi_Q dQ
\end{equation}
and the first law is then satisfied, whatever the value of the two numbers $C$ and $D$. 

That result generalizes to $D\neq0$ those of \cite{Popeetal} where the mass is defined using the Ashtekar-Magon-Das formula, as well as those of \cite{valdivia} which are based on the Hamiltonian approach of \cite{Henneauxetal}, and shows, as emphasized in \cite{Andresetal}, that enforcing that the action be extremum on shell may constrain the integration constants which appear in the general solution of the field equations.

    \section{Concluding remarks}
 
 Solving Einstein's equations to find new black hole solutions has been for decades a huge challenge. Hence, the discovery that only sub-classes of some honest-to-god solutions obeyed the firmly established laws of black hole thermodynamics came as a bad surprise. What was made hopefully translucid in the present paper is that the whole family of solutions is in fact thermodynamically acceptable, under the condition however that a distinction be made between thermodynamical variables and parameters. Indeed the constraint $\phi_2=-3C\phi_1^2+D\ell\phi_1$, see (\ref{phi2condition}), does not numerically constrain $\phi_2$ since $C$ and/or $D$ are arbitrary~; it only tells us that $\phi_2$ cannot be varied independently of $\phi_1$.  It is possible
that these boundary conditions on the scalar field  be
generalized to other families of solutions by modifying the Katz
vector accordingly. For instance, an arbitrary function of the
scalar field times the Katz vector can be considered.

Another conclusion of this paper is that the Einstein-Katz action offers a straightforward way to compute the (automatically finite) black hole action on shell. Moreover, when implementing the constraint on the parameters imposed by the variational principle (that is, $\phi_2=-3C\phi_1^2+D\ell\phi_1$) and using the definitions of the Noether charges as deduced from the KBL superpotential (that is, $M=m_g+D{\phi_1^2/4\ell}$), that action on shell, $I|_{\rm on shell}$, can be related to a Gibbs potential which (automatically) obeys the Gibbs relation (in the case studied here~: $I|_{\rm on shell}=S-\beta(M-Q\Phi_Q)$). That approach has to be contrasted to the usual one where the action is taken to be the Einstein-Hilbert action complemented by the Gibbons-Hawking-York boundary term. Since that action diverges on shell, various counterterms, involving the curvature of the boundary, have to be added, leading to well-controlled and well-understood, albeit fairly heavy calculations, see \cite{Kraus} or \cite{emparan} and \cite{caldarelli} in the case of pure gravity, or \cite{luPopeWen}, \cite{Popeetal} and \cite{Andresetal15} where extra counterterms are added when scalar fields are present.

A question that we leave to future work is how does the Katz boundary action (together with the background bulk action) compare, in general, to the Gibbons-Hawking-York surface term as regularized by the counterterms (which involve curvature tensors). Insight may be gained from \cite{DKconformal} where the KBL vector and superpotential, which, like the GHY term, involve only first derivatives of the metrics, were related to the Ashtekar-Magnon-Das mass formula which, like the counterterms, involves the curvature tensor, that is second derivatives of the metric.

Another question, also left to further studies, is how to generalize the analysis presented here to more general scalar potentials, to higher dimensions, to Gauss-Bonnet or higher derivative theories of gravity.

\begin{acknowledgments}

ND aknowledges very fruitful discussions with Marcela Cardenas and Oscar Fuentealba in Valdivia, with   Olivera Miskovic and Rodrigo Olea in Valparaiso and Paris as well as with Misao Sasaki in Kyoto. She is thankful for financial support provided by CECS in Valdivia, the Pontificia Universidad Cat\'olica de Valpara\`{i}so, and YITP in Kyoto. Research of AA is supported in part by FONDECYT Grants
1141073 and 1161418 and Newton-Picarte Grants DPI20140053 and
DPI20140115.

\end{acknowledgments}


\begin{thebibliography}{99}

\bibitem{Katz85} J. Katz, ``A note on Komar's anomalous factor", Class. Quant. Grav., 2 (1985) 423

\bibitem{KBLB} J. Katz, J.Bicak and D. Lynden-Bell,  ``Relativistic conservation laws and integral constraints for large cosmological perturbations", Phys. Rev. D55 (1997) 5957-5969, see also arXiv: gr-qc/0504041 

\bibitem{Henneauxetal} M. Henneaux, C. Martinez, R. Troncoso and J. Zanelli, ``Asymptotic behavior and Hamiltonian analysis of anti-de Sitter gravity coupled to scalar fields", Annals Phys. 322 (2007) 824 and arXiv: hep-th/0603185

\bibitem{Henningson} M. Henningson and K. Skenderis, ``The holographic Weyl anomaly", JHEP 9807 (1998) 023 and arXiv: hep-th/9806087v2

\bibitem{luPopeWen}H. L\"u, C.N. Pope and Qiang Wen, ``Thermodynamics of AdS Black Holes in Einstein-Scalar Gravity", JHEP 1503 (2015) 165 and arXiv: 1408.1514

\bibitem{Andresetal15} A. Anabal\'on, D. Astefanesei, D. Choque and C. Mart\'inez, ``Trace anomaly and counterterms in designer gravity", JHEP 03 (2016) 117 and arXiv: 1511.08759

\bibitem{Popeetal} H. L\"u, Pang, C.N. Pope, ``AdS Dyonic Black Hole and its Thermodynamics",  JHEP 1311:033, 2013 and arXiv: 1307.6243

\bibitem{Peresetal} A. P\'erez, M. Riquelme, D. Tempo and R. Troncoso,
``Asymptotic structure of the Einstein-Maxwell theory on AdS 3", JHEP 02 (2016) 015 and arXiv: 1512.01576 

\bibitem{Breitenlohner:1982bm}
  P.~Breitenlohner and D.~Z.~Freedman,
  ``Positive Energy in anti-De Sitter Backgrounds and Gauged Extended Supergravity,''
  Phys.\ Lett.\ B  115 (1982) 197


  P.~Breitenlohner and D.~Z.~Freedman,
  ``Stability in Gauged Extended Supergravity,''
  Annals Phys.\   144 (1982) 249
   

\bibitem{Anabalon:2015vda}
  A.~Anabalon, D.~Astefanesei and J.~Oliva,
  ``Hairy Black Hole Stability in AdS, Quantum Mechanics on the Half-Line and Holography,''
  JHEP  1510 (2015) 068 and  arXiv: 1507.05520

\bibitem{JuliaSilva} B. Julia and S. Silva, ``Currents and Superpotentials in classical gauge invariant theories",  Class. Quant. Grav. 15 (1998) 2173-2215  and arXiv: gr-qc/9804029


  
\bibitem{DKO} N. Deruelle, J. Katz, S. Ogushi, ``Conserved charges in Einstein Gauss-Bonnet theory", Class. Quant. Grav. 21 (2004) 1971-1985 and arXiv: gr-qc/0310098

 \bibitem{DKkerr} N. Deruelle and J. Katz, ``On the mass of a Kerr-anti-de Sitter spacetime in D dimensions", Class. Quant. Grav. 22 (2005) 421-424 and arXiv:  gr-qc/0410135
 
 \bibitem{DM} N. Deruelle and Y. Morisawa, ``Mass and angular momenta of Kerr anti-de Sitter spacetimes in Einstein-Gauss-Bonnet theory",  Class. Quant. Grav. 22 (2005) 933-938 and arXiv:  gr-qc/0411135
 
 \bibitem{Petrov} A. N. Petrov, ``Noether and Belinfante corrected types of currents for perturbations in the Einstein-Gauss-Bonnet gravity", Class. Quant. Grav. 28 (2011) 215021 and  arXiv: 1102.5636 
 
\bibitem{DKconformal} N. Deruelle and J. Katz, ``Comments on conformal masses, asymptotics backgrounds and conservation laws",  Class. Quant. Grav. 23 (2006) 753-760, arXiv: gr-qc/0512077 

\bibitem{Klemm} F. Faedo, D. Klemm, and M. Nozawa ``Hairy black holes in N = 2 gauged supergravity", JHEP 1511 (2015) 045 and arXiv: 1505.02986


\bibitem{LiuLu} Hai-Shan Liu and H. L\"u. ``Scalar Charges in Asymptotic AdS Geometries", Phys. Lett. B730 (2014) 267-270 and arXiv:1401.0010 

\bibitem{Andresetal} A. Anabal\'on, D. Astefanesei and C. Mart\'\i nez, ``Mass of asymptotically anti-de Sitter hairy spacetimes", Phys. Rev. D91, 041501 (2015) and arXiv:1407.3296

\bibitem{Andressol} A. Anabal\'on, ``Exact Black Holes and Universality in the Backreaction of non-linear Sigma Models with a potential in (A)dS4", JHEP 06 (2012) 127 and arXiv: 1204.2720

\bibitem{valdivia} M. C\'ardenas, O. Fuentealba and J. Matulich,  ``A note on the AdS4 dyonic black hole thermodynamics", JHEP 05 (2016) 001 and arXiv: 1603.03760

\bibitem{Compere} D.D.K. Chow and G. Compere, ``Dyonic AdS black holes in maximal gauged supergravity", Phys. Rev. D 89, 065003 (2014) and arXiv: 1311.1204 

\bibitem{Kraus} V. Balasubramanian and P. Kraus, ``A Stress tensor for Anti-de Sitter gravity", Commun. Math. Phys. 208 (1999) 413  and arXiv: hep-th/9902121

\bibitem{emparan}  R. Emparan, C. V. Johnson, and R. C. Myers, ``Surface terms as counterterms in the AdS/CFT correspondence," Phys. Rev. D60 (1999) 104001 and arXiv: hep-th/9903238

\bibitem{caldarelli} M.M. Caldarelli, G. Cognola and D. Klemm, ``Thermodynamics of Kerr-Newman-AdS Black Holes and Conformal Field Theories", Class. Quant. Grav. 17 (2000) 399-420 and arXiv: hep-th/9908022

\end{thebibliography}
\end{document}